# Forecasting and Analyzing the Military Expenditure of India Using Box-Jenkins ARIMA Model

Deepanshu Sharma
Amity School of Engineering & Technology,
GGS Indraprastha University,
Delhi, India,
sharma.deepanshu97@gmail.com

Kritika Phulli
Amity School of Engineering & Technology,
GGS Indraprastha University,
Delhi, India,
kritika1298phulli@gmail.com

**Abstract:** The advancement in the field of statistical methodologies to economic data has paved its path towards the dire need for designing efficient military management policies. India is ranked as the third largest country in terms of military spender for the year 2019. Therefore, this study aims at utilizing the Box-Jenkins ARIMA model for time series forecasting of the military expenditure of India in forthcoming times. The model was generated on the SIPRI dataset of Indian military expenditure of 60 years from the year 1960 to 2019. The trend was analysed for the generation of the model that best fitted the forecasting. The study highlights the minimum AIC value and involves ADF testing (Augmented Dickey-Fuller) to transform expenditure data into stationary form for model generation. It also focused on plotting the residual error distribution for efficient forecasting. This research proposed an ARIMA (0,1,6) model for optimal forecasting of military expenditure of India with an accuracy of 95.7%. The model, thus, acts as a Moving Average (MA) model and predicts the steady-state exponential growth of 36.94% in military expenditure of India by 2024.

## 1. Introduction

The Indian military, with a strength of over 1.4 million active personnel [1,] [2], is believed to be the world's second largest military force [3]. India, which has been characterized as the world's largest democracy, happens to be the pre-eminent and emerging developing country (EDC) concerning military expenditure and military manufacturing. As per the reports of Stockholm International Peace Research Institute (SIPRI) for the year 2019, India has become the third largest military expender in the world following US and China, accounting to the expenditure of $71.1 billion [4]. Comparing this record to the former year (2018), India moved from fourth position to third position, surpassing Saudi Arabia. Subsequently, the Indian military forces have been ranked 4[th] country among 138 countries for the annual GFP (Global Fire Power) review with an index of 0.0953 [22]. While exploring the SIPRI reports for military expenditure, a growth of 259% has been observed in the past 30 years [5]. Contemplating these facts, it has become a needful piece of work to predetermine the forthcoming expenditure on military for planning the strategies efficiently in advance. This can be implemented using the time series forecasting approach. Time series forecasting, unlike other popular AI approaches, is an endeavour of extracting meaning information from the historic data and utilizing it from predicting the future events based on the time component. These problems can be considered supervised learning problems but they have to be retrained every time a new prediction has to be generated. Models such as decomposition models, smoothing models, ARIMA

models, exponential models etc. can be used in time series forecasting. This study uses a quantitative method, ARIMA, for time series forecasting as it only requires the prior data to generalize the forecast and also increases the accuracy of forecast while keeping the number of parameters minimum. Autoregressive integrated moving average (ARIMA) model are statistically sophisticated and mathematically complex methods, which best represents a time series by modelling the correlations in the data. ARIMA utilizes box- Jenkins methodology [6] combines the individual autoregressive (AR) and moving average (MA) techniques.

Taking into account the above applications, a similar attempt has been made in this study to predict the military expenditure in India for the forthcoming years, considered over a time span of 60 years till 2019 using the ARIMA model of time series forecasting. This dataset has been derived from the World Bank Data Catalogue and has been indicated in the SIPRI Military Expenditure Database, 2019 [7] having consistent time series [12]. The expenditure is expressed as a share of GDP in US $ according to the calendar year. This forecasting would help at analysing the military expenditures ahead of time. Recently, a similar study was conducted for analyzing the China's military expenditure [8]. Tensions in India's relations with its immediate neighbour-China makes it a dire requirement for this research to be conducted so that it facilitates the formulation of effective workplans for permitting military budget and its expenditure. The research would also help in analysing the gap between military demand and expenditure by government of India to better formulate policies leaping to minimize the gap and pacing toward a military might status in the world. Many researchers have actively dedicated their studies in designing an optimised model for forecasting using ARIMA [9], [10], [11]. However, little amount of research has been performed for the prediction of the military expenditure of India using Box-Jenkins ARIMA approach. Thus, Box- Jenkins time series ARIMA model was preferred to forecast the military expenditure of India over the forthcoming 5 years.

*Research Contributions*

1. Analyzed the time series trend of the military expenditure data for 60 years.
2. Determined the minimum AIC score of the model and validated through the Augmented Dickey fuller (ADF) test.
3. Predicted the military expenditure for over 5 years with an accuracy of 95.7%
4. Plotted residual error distribution for efficient forecasting.
5. Forecasted a 36.94% increase in military expenditure of India by 2024.

## 2. Related Work

The global military expenditure for the year 2019 has been estimated to $1917 billion as per the report of Trends in World Expenditure, 2019 [4]. In 2019, India contributed around 3.7% of the entire world's expenditure and stands at the third position after US and China. Considering the past records, there has been an increase in the spending trend since the previous years due to the increasing conflicts between its neighbouring countries. In the past, studies have analyzed the military expenditure of countries like Russia concluding that modernization of the Armed Forces continues to be a high priority. [21] However, the domain of military expenditure in India has been less explored till now.

Additionally, over the past years, there have been many researches in the field of time series forecasting modelling [8], [9], [11], [18], [20]. Time series predication plays a significant role for future planning and advances in various field of study such as stock market, consumer demands etc [18]. ARIMA model, being the premium time series forecasting model has been applied widely for water quality prediction [9], consumer's expenditure [11], energy price [10]. Though, there stands a condition for stationary data for ARIMA modelling [17].

## 3. Research Methodology

This research forecasts the India's military expenditure for forthcoming 5 years using the quantitative time series forecasting model, ARIMA. The dataset was obtained from the SIPRI Military Expenditure Database, 2019 for the timestamp of 60 years with yearly expenditure in US $ in billions considering the value of $ in early 2020s. It is worth noting that Arima models are generated for forecasting time series military expenditure data in this study. The data initially was analysed for observing the trend in the raw data. While forecasting using ARIMA model, Stationarity and differencing has to be taken into consideration. Once the conditions are satisfied and validated using Augmented Dicker Fuller test, the ARIMA model was generated using the 3 parameters: p, q and d. The parameters were evaluated through the order of their respective functions explained in later section. The workflow of the study is described in fig. 1.

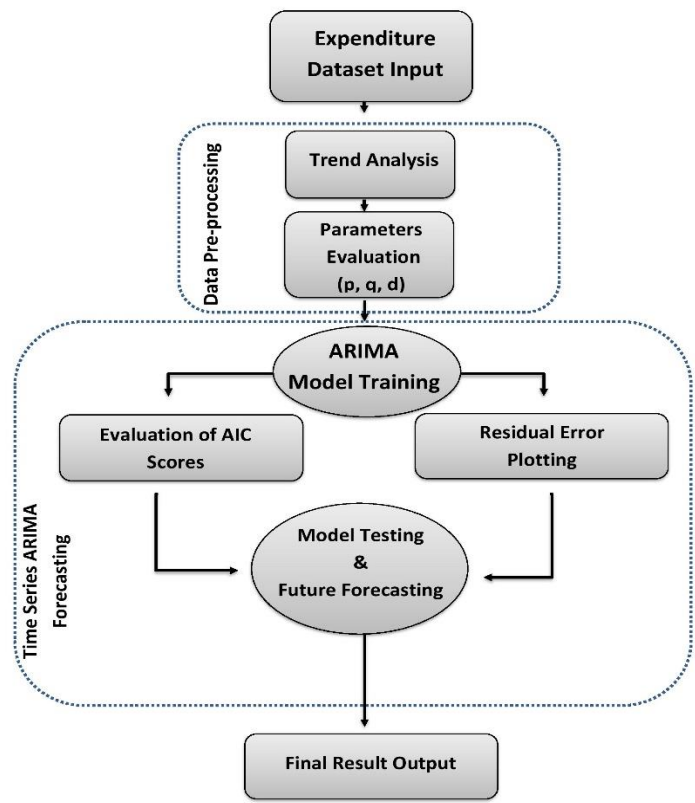

*Fig. 1: Workflow of the Study*

## 3.1. The Proposed Model: Auto Regressive Integrated Moving Average Model

The ARIMA Model is an integrated approach of the two models: autoregressive (AR) model, moving average (MA) model and seasonal ARIMA (SARIMA) [13], [15], [17].

• Auto Regression refers to the model that predicts the output results based on a linear combination of historic values of the variable [14]. This model uses the dependent relationship between the observation and some number of lagged observations. Therefore, the AR(p) model of order p can be represented as:

$$y_t = c + \varphi_1 y_{t-1} + \varphi_2 y_{t-2} + \cdots + \varphi_p y_{t-p} + \varepsilon_t \qquad (1)$$

Where,

$\varepsilon_t$ = white noise (randomness)

P= order (1,2, 3...)

$y_{t-1}, y_{t-2}, y_{t-p}$ = past series values (lags)

$c = \left(1 - \sum_{i=1}^{p} \phi_i \right) \mu$      where $\mu$ = process mean

• Moving Average model uses past forecast errors in regression- like model [16]. This model uses the dependency between an observation and a residual error from a moving average model applied to lagged observations. The MA(q) model is defined as:

$$y_t = c + \varepsilon_t + \varphi_1 \varepsilon_{t-1} + \varphi_2 \varepsilon_{t-2} + \cdots + \varphi_p \varepsilon_{t-q} \qquad (2)$$

Where,

$y_t$ = weighted moving average of past few forecast errors

$\varepsilon_t$ = white noise (randomness)

q= order (1,2, 3...)

• The term 'Integrated' refers to the use of differencing of raw observations for making the time series stationary. Thus, a linear regression model is constructed using the ARIMA method and is denoted by ARIMA (p, d, q) where each specified parameter has an integer value depicting their respective nature.

- p – The number of lag observations ~ lag order.
- d – The number of times the raw observations are differenced ~ order of differencing
- q – The size of the moving average window ~ order of moving average.

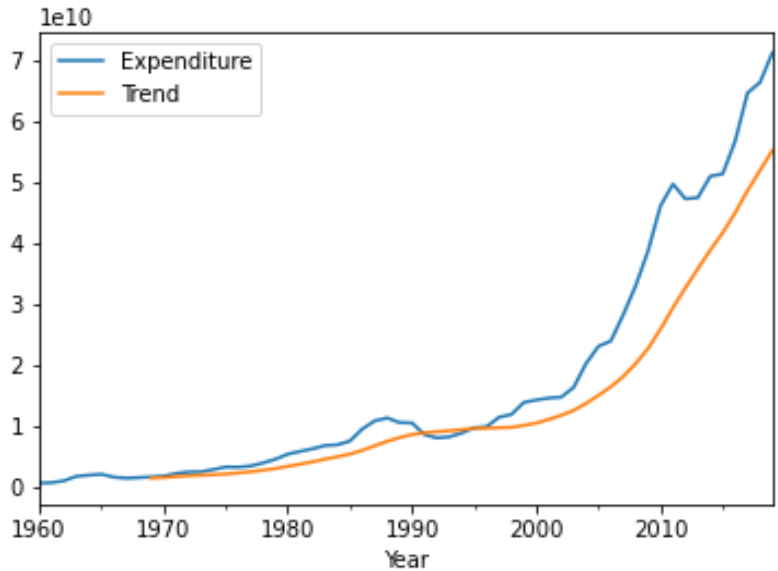

*Fig. 2: Trend of Military Expenditure Observed over 60 years*

The listed below is the step by step procedure followed for generating an efficient ARIMA based forecasting model. It is important to note that all the data visualization and model generation is implemented through Python framework and its libraries.

*Step 0: Analysis of the Trend:* This is the initial visualization of the data before any pre-processing. Fig. 2 is the pictorial representation of the trend of the military expenditure of India from the year 1960-2019.

*Step 1: Model Identification:* Data must be verified as stationary for efficient results. This requirement is fulfilled through Augmented Dickey Fuller (ADF) test. If the test fails (data: non- stationary), it is required to make it stationary using the differencing methodology.

Hypothesis for ADF test.

$H_0$: A unit root is present in a time series sample.

$H_1$: A unit root is stationarity or trend stationarity.

The equation for ADF testing:

$$\triangle y_t = \alpha + \beta t + \gamma y_{t-1} + \delta_1 \triangle y_{t-1} + \cdots + \delta_{p-1} \triangle y_{t-p+1} + \varepsilon_t \quad (3)$$

Where,

α=constant;

β = coefficient of time trend;

p = lag order of auto regressive process

Table 1: Augmented Dickey Fuller (ADF) Test Observations

| Data Point | p-Value | Critical Value | | |
|---|---|---|---|---|
| | | 1% | 5% | 10% |
| Non - Transformed | 0.99 | -3.568 | -2.921 | -2.598 |
| Log -Transformed | 0.0001 | -3.48 | -2.912 | -2.594 |

*Initially, the dataset was log transformed to stabilize the data. After log transforming the value of d=1 (first order differentials) was utilised as the differencing parameter to make the series of military expenditure stationary. The ADF statistic is a negative number. More negative, Stronger to reject the hypothesis that there is a unit root at some level of confidence. Table 2 depicts the results of ADF test before and after transformation.*

<u>Step2: Parameter Estimation</u> – For ARIMA process, the data should have autocorrelation properties. The correlation is discovered by the use of Auto correlation Function (ACF) and Partial Auto Correlation Function (PACF) for identification of the order of ARIMA model (parameters p and q) [19]. The optimal model is selected based on the minimum Akaike Information Criteria (AIC) value. Subsequently, the minima point of residual is achieved through a non-linear optimization process. The residuals describe the goodness of fit of model meeting the requirement of the white noise process.

The formula of AIC:

$$AIC = -2\log(L) + 2k \qquad (4)$$

Where,

L= Likelihood of the data

K= number of parameter (including $\sigma^2$)

*In the above steps, using the value of d = 1, the series is then analysed for the ACF and PACF functions. The ACF and PACF graphs are plotted and represented in the fig. 3. The ACF graph is used to determine q value and the PACF graph determines the p values based on the results of differencing.*

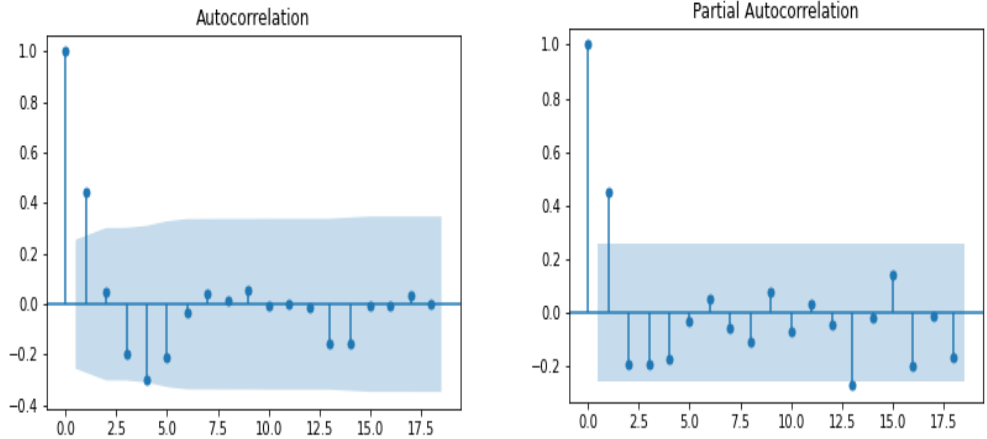

*Fig. 3: Autocorrelation (ACF) and Partial Autocorrelation (PACF) plots of Log Transformed data*

*Step3:* Training and Testing the Model: Here, the dataset is divided into 7:3 for training and testing purposes and then fitted to the model using the parameters ARIMA (p, d, q). For achieving the stability for the forecast, this study uses a range of p and q values so that a minimum AIC score is achieved and the best accuracy among the considered are extracted. The following observations are represented in Table 2.

*Table. 2: Akaike Information Criteria (AIC) Scores for various ARIMA models*

| ARIMA (p, d, q) | AIC Score |
|---|---|
| (0,1,5) | -74.89 |
| **(0,1,6)** | **-76.02** |
| (2,1,2) | -73.06 |
| (2,1,5) | -73.89 |
| (3,1,0) | -73.94 |

The Python library used for modelling ARIMA:

**from *statsmodels.tsa.arima_model import ARIMA***

*Since the minimum AIC score is depicted by p=0, q=1, d=6. Hence, ARIMA (0,1,6) is modelled. Using these parameters, the dataset is trained and tested. Fig. 4 illustrates the predicted vs actual plot for the considered ARIMA Model.*

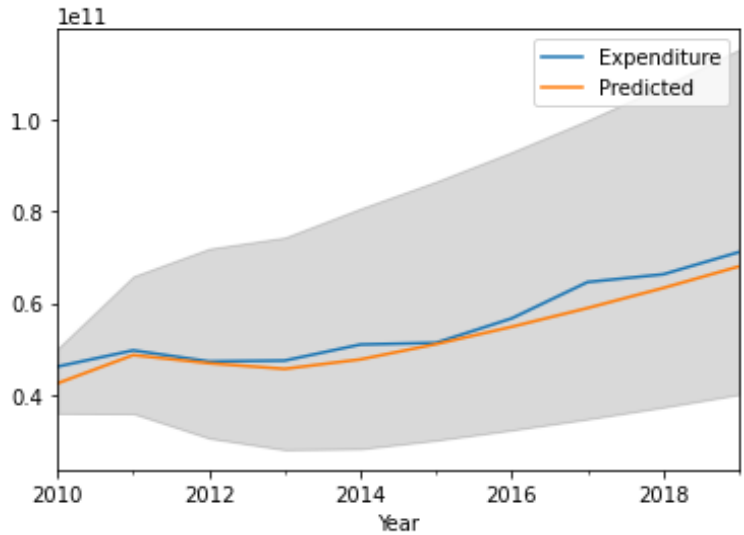

*Fig. 4: Predicted vs Actual Plot for ARIMA (0,1,6)*

*Step4: Diagnostic check*: This step determines for the residuals of ARIMA models obtained from ACF and PACF graphs to be independent and identically distributed. It testifies whether they satisfy the characteristics of a white noise process. The residual error plots of the model are depicted in the fig. 5. Also, Table 3 and Table 4 illustrates the model parameters and root values of the generated ARIMA Model.

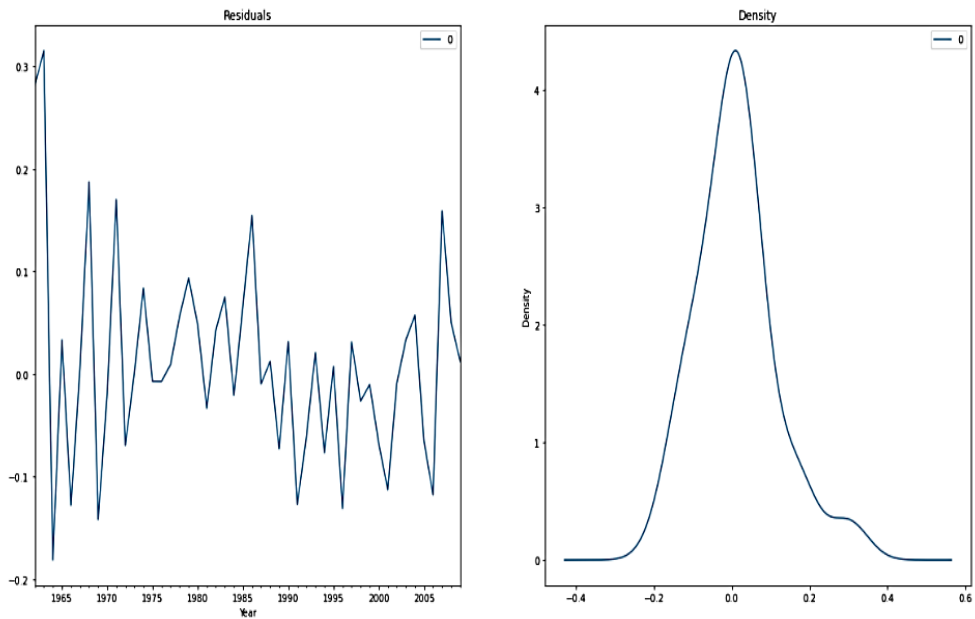

*Fig. 5: Residuals Error Plots of ARIMA (0,1,6) Model*

*Table 3: ARIMA (0,1,6) Model Parameters*

| Dependent Variable: | D.newLog | No. Observations: | 48 | | | |
|---|---|---|---|---|---|---|
| Model: | ARIMA(0, 1, 6) | Log Likelihood | 46.012 | | | |
| | | S.D. of innovations | 0.085 | | | |
| | | AIC | -76.025 | | | |
| | | BIC | -61.055 | | | |
| | | HQIC | -70.368 | | | |
| | Coefficient | std err | z | P>|z| | [0.025 | 0.975] |
| Constant | 0.0719 | 0.006 | 12.94 | 0 | 0.061 | 0.083 |
| MA.L1.D.newLog | 0.5263 | 0.171 | 3.087 | 0.002 | 0.192 | 0.861 |
| MA.L2.D.newLog | 0.2898 | 0.176 | 1.65 | 0.099 | -0.054 | 0.634 |
| MA.L3.D.newLog | -0.4071 | 0.191 | -2.126 | 0.033 | -0.782 | -0.032 |
| MA.L4.D.newLog | -0.2423 | 0.159 | -1.523 | 0.128 | -0.554 | 0.07 |
| MA.L5.D.newLog | -0.8283 | 0.17 | -4.886 | 0 | -1.16 | -0.496 |
| MA.L6.D.newLog | -0.3379 | 0.190 | -1.775 | 0.076 | -0.711 | 0.035 |

*Table 4: Root Values of Moving Average in ARIMA (0,1,6) Model*

| | Real | Imaginary | Modulus | Frequency |
|---|---|---|---|---|
| MA.1 | 1.0001 | -0.0000j | 1.0001 | 0 |
| MA.2 | 0.197 | -1.0899j | 1.1076 | -0.2215 |
| MA.3 | 0.197 | +1.0899j | 1.1076 | 0.2215 |
| MA.4 | -0.7164 | -0.6977j | 1 | -0.3771 |
| MA.5 | -0.7164 | +0.6977j | 1 | 0.3771 |
| MA.6 | -2.4125 | -0.0000j | 2.4125 | -0.5 |

*Step5: Forecasting the data:* Using the model selected from above Diagnostic check, the future military expenditure was forecasted for forthcoming years along with its lower bound and upper bound values with 95% confidence. The forecasted values with their lower and upper bounds are depicted in Table 5.

*Table 5: Forecasted Values for forthcoming years along with Lower and Upper Bound in US billion $*

| Year | Forecasted (US $ in billion) | Lower Bound (US $ in billion) | Upper Bound (US $ in billion) |
|---|---|---|---|
| 2020 | 73.04 | 43.01 | 124.05 |
| 2021 | 78.49 | 46.21 | 133.3 |
| 2022 | 84.34 | 49.66 | 143.23 |
| 2023 | 90.62 | 53.36 | 153.9 |
| 2024 | 97.37 | 57.34 | 165.37 |

## 4. Results and Discussions

This section summarizes the findings of the research. As mentioned in the above sections, the first and foremost outcomes depicts the trend of the time series data. Fig. 2. illustrates an exponential increasing trend of the military expenditure.

Once the trend has been analysed, the data was transformed into its stationary form for modelling. For this, the data was pre-processed using the log transformation and differencing value (shift=1) and was validated using ADF test. Before transformation of data, ADF failed but after log transformation, ADF test accepted the null hypothesis, approximating to 0.0001. The results of ADF test are reflected in Table 1.

Using the differencing value (d=1) where the data is validated to be stationary, the other parameters: p and q, are estimated using the ACF and PACF graphs. The graphs are depicted in the fig. 3.

The dataset was then used for training and testing purposes, divided into the ratio of 7:3 for generated the ARIMA model. A range of p and q values were used for modelling to achieve the stability of the generated model. For the range of p and q values, a respective AIC score is computed which is depicted in Table 2. The minimum AIC score was considered to select the optimal p and q values for further evaluations. Parameters obtained for optimal modelling of the time series data using ARIMA method are as follows:

$$p: 0; \quad d: 1 \quad q: 6;$$

The AIC value came out to be -76.03, which is minimum among the others.

It is worth mentioning that the actual and predicted values of testing data comes out to be approximately close to each other with an accuracy of 95.73%. The plot is described in fig. 4. The residuals of the model after fitting are depicted in fig. 5.

The parameters of the generated model are elaborated in Table 3 along with their root values of Moving Average model which is depicted in Table 4. After the model generation for ARIMA (0,1,6), the forecasted values for the forthcoming years (2020-2024) are illustrated in the Table 5.

*This ARIMA Model generated on p=0, d=1, q=6; having AIC value = -76.03 acts as a Moving Average (MA) Model with an accuracy of 95.73%.*

## 5. Conclusion and Future Scope

The Indian military forces have been designated as one of the top five powers in the world and has also been ranked 4th out of 138 countries for the annual GFP (Global Fire Power) review with an index of 0.0953 [22]. Also, India has become the third largest military expender in the world following US and China, accounting to the expenditure of $71.1 billion [4] for the year 2019. Since the emerging tensions among its neighbouring countries, India needs to be a superpower for battling with all its well-equipped resources. For buying and manufacturing powerful military equipment, it is a basic necessity to realize the military expenditure for the forthcoming years for well-structured and innovative planning. For the same purpose, the proposed model in this study analyses the 60 years of annual military expenditure of India using the ARIMA technique of time series forecasting. The dataset was collected from the SIPRI Military Expenditure Database, 2019. The military expenditure is considered in US billion $ value for the early 2020s. The proposed ARIMA model attains an accuracy of 95.73% with an increasing trend and a differencing value of 1. The estimated parameters for auto regressive (AR) and moving average (MA) are p= 0 and q= 6 respectively using the PACF and ACF plotted graphs. The AIC score acquired accounts to -76.03 with the above formulated parameters, describing a better fitted ARIMA model when compared to other values of the parameters. This generated model acts as a Moving Average (MA) model since p value comes out to be 0. This proposed model not only helps to forecast the military expenditure for the future but would also help in allocation of the military budget for substantial growth of the country.